\documentstyle[preprint,prc,aps,floats,epsf]{revtex}

\tightenlines

\begin{document}

%
%
\def\symdef#1#2{\def#1{#2}}
%

\symdef{\alphabar}{\overline\alpha}
\symdef{\alphabarp}{\overline\alpha\,{}'}
\symdef{\Azero}{A_0}

\symdef{\bzero}{b_0}

\symdef{\Cs}{C_{\rm s}}
\symdef{\cs}{c_{\rm s}}
\symdef{\Cv}{C_{\rm v}}
\symdef{\cv}{c_{\rm v}}

\symdef{\dalem}{\frame{\phantom{\rule{8pt}{8pt}}}}
\symdef{\del}{\partial}
\symdef{\delN}{{\wt\partial}}
\symdef{\Deltaevac}{\Delta{\cal E}_{\rm vac}}

\symdef{\ed}{{\cal E}}
\symdef{\edk}{{\cal E}_k}
\symdef{\edkzero}{{\cal E}_{k0}}
\symdef{\edv}{{\cal E}_{\rm v}}
\symdef{\edvphi}{{\cal E}_{{\rm v}\Phi}}
\symdef{\edvphizero}{{\cal E}_{{\rm v}\Phi 0}}
\symdef{\edzero}{{\cal E}_{0}}
\symdef{\Efermistar}{E_{{\scriptscriptstyle \rm F}}^\ast}
\symdef{\Efermistarzero}{E_{{\scriptscriptstyle \rm F}0}^\ast}
\symdef{\etabar}{\overline\eta}
\symdef{\ezero}{e_0}

\symdef{\fomega}{f_\omegav}
\symdef{\fpi}{f_\pi}
\symdef{\fv}{f_{\rm v}}
\symdef{\fvt}{\widetilde\fv}

\symdef{\gA}{g_A}
\symdef{\gammazero}{\gamma_0}
\symdef{\gomega}{g_\omegav}
\symdef{\gpi}{g_\pi}
\symdef{\grho}{g_\rho}
\symdef{\grad}{{\bbox{\nabla}}}
\symdef{\gs}{g_{\rm s}}
\symdef{\gv}{g_{\rm v}}

\symdef{\fm}{\mbox{\,fm}}

\symdef{\infm}{\mbox{\,fm$^{-1}$}}
\symdef{\isovectorTensor}{s_{\tauvec}}
\symdef{\isovectorTensorN}{\wt\isovectorTensor}
\symdef{\isovectorVector}{j_{\tauvec}}
\symdef{\isovectorVectorN}{\wt j_{\tauvec}}

\symdef{\kappabar}{\overline\kappa}
\symdef{\kfermi}{k_{{\scriptscriptstyle \rm F}}}
\symdef{\kfermizero}{k_{{\scriptscriptstyle \rm F}0}}
\symdef{\Kzero}{K_0}

\symdef{\lambdabar}{\overline\lambda}
\symdef{\LdotS}{\bbox{\sigma\cdot L}}
\symdef{\lsim}{\lower0.6ex\vbox{\hbox{$\ \buildrel{\textstyle <}
         \over{\sim}\ $}}}
\symdef{\lzero}{l_{0}}

\symdef{\Mbar}{\overline M}
\symdef{\Mbarzero}{\Mbar_0}
\symdef{\MeV}{\mbox{\,MeV}}
\symdef{\momega}{m_\omegav}
\symdef{\mpi}{m_\pi}
\symdef{\mrho}{m_\rho}
\symdef{\ms}{m_{\rm s}}
\symdef{\Mstar}{M^\ast}
\symdef{\Mstarzero}{M^\ast_0}
\symdef{\mv}{m_{\rm v}}
\symdef{\mzero}{{\rm v}_{0}}

\symdef{\Nbar}{\overline N}

\symdef{\omegaV}{V}
\symdef{\omegav}{{\rm v}}

\symdef{\Phizero}{\Phi_0}
\symdef{\psibar}{\overline\psi}
\symdef{\psidagger}{\psi^\dagger}
\symdef{\pvec}{{\bf p}}

\symdef{\rhoB}{\rho_{{\scriptscriptstyle \rm B}}}
\symdef{\rhoBt}{\wt\rho_{{\scriptscriptstyle \rm B}}}
\symdef{\rhoBzero}{\rho_{{\scriptscriptstyle \rm B}0}}
\symdef{\rhominus}{\rho_{-}}
\symdef{\rhoplus}{\rho_{+}}
\symdef{\rhos}{\rho_{{\scriptstyle \rm s}}}
\symdef{\rhospzero}{\rho'_{{\scriptstyle {\rm s} 0}}}
\symdef{\rhost}{\wt\rho_{{\scriptstyle \rm s}}}
\symdef{\rhoszero}{\rho_{{\scriptstyle {\rm s}0}}}
\symdef{\rhotau}{\rho_{\tauvec}}
\symdef{\rhotaut}{\wt\rho_{\tauvec}}
\symdef{\rhothree}{\rho_{3}}
\symdef{\rhothreet}{\wt\rho_{3}}
\symdef{\rhozero}{\rho_0}

\symdef{\scalar}{\rhos}
\symdef{\scalarN}{{\rhost}}
\symdef{\Szero}{S_0}

\symdef{\tauvec}{{\bbox{\tau}}}
\symdef{\tauthree}{\tau_3}
\symdef{\tensor}{{{s}}}
\symdef{\tensorN}{\wt\tensor}
\symdef{\tensort}{\wt{{s}}}
\symdef{\tensorthree}{\tensor_3}
\symdef{\tensorthreet}{\wt\tensorthree}
\symdef{\Tr}{{\rm Tr\,}}

\symdef{\umu}{u^\mu}
\symdef{\Ualpha}{U_{\alpha}}
\symdef{\Ueff}{U_{\rm eff}}
\symdef{\Uzero}{U_0}
\symdef{\Uzerop}{U_0'}
\symdef{\Uzeropp}{U_0''}

\symdef{\vecalpha}{{\bbox{\alpha}}}
\symdef{\veccdot}{{\bbox{\cdot}}}
\symdef{\vecnabla}{{\bbox{\nabla}}}
\symdef{\vecpi}{{\bbox{\pi}}}
\symdef{\vectau}{{\bbox{\tau}}}
\symdef{\vector}{j_{\scriptscriptstyle V}}
\symdef{\vectorN}{{\wt\vector}}
\symdef{\vecx}{{\bf x}}
\symdef{\Vopt}{V_{\rm opt}}
\symdef{\Vzero}{V_0}

\symdef{\wt}{\widetilde}
\symdef{\wzero}{w_0}
\symdef{\Wzero}{W_0}

\symdef{\zetabar}{\overline\zeta}
%
%
%
%

\def\beq{\begin{equation}}
\def\eeq{\end{equation}}
\def\beqa{\begin{eqnarray}}
\def\eeqa{\end{eqnarray}}
\def\smk{\kern2pt}

\draft

\preprint{IU/NTC\ \ 99--10}

\title{Parameter Counting in Relativistic\\
       Mean-Field Models}

\author{R. J. Furnstahl and Brian D. Serot\thanks{%
        Permanent address: Department of Physics and Nuclear Theory Center,
         Indiana University,\ \ Bloomington, IN\ \ 47405;
		 serot@iucf.indiana.edu.}}

\address{Department of Physics \\
         The Ohio State University,\ \ Columbus, OH\ \ 43210}
%
%
%

%
\date{October, 1999}
\maketitle
\begin{abstract}
Power counting is applied
to  relativistic mean-field energy functionals to estimate
contributions to the energy from individual terms.
New estimates for isovector, tensor, and gradient terms in finite
nuclei are shown to be consistent with direct, high-quality fits.
The estimates establish a hierarchy of model parameters and identify how many
parameters are well constrained by
bulk nuclear observables.
We conclude that four (possibly five) isoscalar, non-gradient parameters, 
one gradient parameter, and one isovector parameter
are well determined by the usual bulk nuclear observables.
\end{abstract}

\smallskip

\section{Introduction and Motivation}

Relativistic mean-field (RMF) descriptions of the nuclear many-body problem  
have existed for nearly fifty years \cite{SCHIFF51}.
An extensive body of work  \cite{SEROT86,SEROT92,SEROT97} has shown that RMF
theories can realistically reproduce the bulk and single-particle
properties of medium to heavy nuclei.
Successful RMF theories are characterized by large, 
neutral,
Lorentz scalar and vector potentials (roughly several hundred MeV at nuclear
equilibrium density), a new saturation mechanism for nuclear matter that
arises from relativistic ({\it i.e.\/}, 
velocity-dependent) interaction effects from the scalar
potential, and a nuclear spin-orbit force that is determined by this velocity
dependence.
The vector potential provides an efficient representation of the
short-range nucleon--nucleon (NN) repulsion, while the scalar potential 
provides the mid-range NN attraction by 
simulating correlated, scalar-isoscalar two-pion exchange, which is
the most important pionic contribution
for describing the bulk properties of nuclear matter.

For many years, RMF model parameters were determined by fitting them
to the properties of nuclei or nuclear matter using various ``black-box''
fitting schemes \cite{HOROWITZ81,REINHARD86,RUFA88}.
Direct connections between the RMF parameters and the systematics of nuclear
observables have begun to emerge only recently 
\cite{BODMER91,FURNSTAHL93,FURNSTAHL96a,FURNSTAHL97}.
The modern theoretical viewpoint underlying the RMF, which relies
on the ideas of Effective Field Theory (EFT) and of Density Functional Theory
(DFT) \cite{FURNSTAHL97,SEROT97,RUSNAK97}, 
provides a systematic truncation procedure for the RMF energy
functional.
However, it introduces more parameters than can be unambiguously
determined by the relevant data.
The present work uses EFT power counting
to examine the quantity of 
information contained in the bulk and single-particle input data, 
and to identify how
many parameters can be reliably determined by these data.

The original Walecka model 
\cite{WALECKA74}
({\it without\/} nonlinear field interactions), 
contained two free isoscalar parameters
in nuclear matter, which were used to fit the equilibrium density and
binding energy.
An isovector coupling was added \cite{SEROT79} to reproduce the bulk symmetry 
energy, and for finite nuclei, a single scale parameter (the scalar ``meson''
mass) was chosen to reproduce the rms charge radius of ${}^{40}$Ca.
These procedures yielded reasonable results for charge-density distributions
of spherical nuclei and for single-particle spectra, revealing that the
nuclear shell model was obtained automatically
without any specific adjustment of 
spin-orbit parameters \cite{MILLER74,BROCKMANN78,SAVUSHKIN79,HOROWITZ81}.
Nevertheless, the nuclear compressibility and surface energy were too large
in these simple calculations, leading to incorrect systematics in the 
total binding energies.
Moreover, no quantitative measure of the accuracy of the results was obtained.

It was also known at that time that the compressibility could be reduced by
adding cubic and quartic scalar self-couplings to the model \cite{BOGUTA77}.
This procedure established a basic phenomenology for RMF studies of nuclei
that is still used widely 
\cite{BOUYSSY84,REINHARD86,FURNSTAHL87,RUFA88,REINHARD89,GAMBHIR90,SHARMA94,MAHARANA96}.
Quantitative measures of the accuracy of the results were
introduced, showing that there was a significant improvement over the
original (linear) model.
However, attempts to further improve the accuracy met with little success;
for example,
the introduction of a quartic vector self-coupling \cite{GMUCA92}
allowed one to perform RMF calculations of nuclei based on 
Dirac--Brueckner--Hartree--Fock calculations of infinite nuclear matter,
but little improvement, if any, was achieved in reproducing the bulk and
single-particle nuclear properties.

An important advance was to 
{\it define\/} a measure of accuracy for the fit and then {\it search\/}
the parameter space (in some particular model) to find a set of ``optimal''
parameters \cite{RUFA88,REINHARD89,SHARMA94}.
The  conclusion of these surveys was that the linear Walecka
model plus cubic and quartic scalar self-couplings exhausted the
capabilities of mean-field phenomenology to reproduce the relevant nuclear 
data \cite{RUFA88}.
However, the associated mean-field energy functional was 
truncated solely on the basis of its empirical success,
essentially without theoretical justification.
Our basic goal is to reassess this issue using the modern EFT framework
for the nuclear many-body problem\cite{FURNSTAHL97,PANIC99} 
to ascertain how many model
parameters (or linear combinations thereof) are accurately determined
by input data drawn from bulk and single-particle nuclear properties.

\section{Energy Functional and Power Counting Estimates}

Although we could use meson models as done historically, the
analysis is more transparent with ``point-coupling'' 
models \cite{NIKOLAUS97,FRIAR96,RUSNAK97}, 
which contain only nucleon fields in a local lagrangian.
Because of the freedom to perform field
redefinitions, a general point-coupling model is equivalent to
a general meson model \cite{FURNSTAHL97,RUSNAK97}.
An energy functional of nucleon densities can be
constructed by starting with
a general point-coupling effective lagrangian,  
consistent with
the underlying symmetries of QCD ({\it e.g.}, Lorentz covariance,
electromagnetic gauge invariance, and spontaneously broken chiral symmetry),
and by constructing the corresponding one-loop 
energy functional.
As discussed in Ref.~\cite{SEROT97},
this approach approximates a general DFT functional that
incorporates many-body effects beyond the Hartree 
level 
when parameters are determined from finite-density data \cite{DREIZLER90}.
We expect this factorized representation of the functional 
to be reasonable because the large
scalar and vector mean-field potentials 
imply that products of field operators in the lagrangian
will be well approximated by products of densities in the functional
(``Hartree dominance'') \cite{FURNSTAHL97,SEROT97}. 
More sophisticated approximations to the many-body problem
({\it e.g.}, Dirac--Hartree--Fock, Dirac--Brueckner--Hartree--Fock, 
etc.\ \cite{BOUYSSY87,BLUNDEN88,ANASTASIO83,HOROWITZ87,BROCKMANN84,MALFLIET91})
could lead to more complete energy functionals, but these calculations are not
relevant for the questions studied here.

The DFT framework generates a functional with an unlimited number of terms
and corresponding parameters.
It is useful only if we can identify a valid
expansion and truncation scheme.
This requires an organization of terms in the effective lagrangian
and a way to estimate the couplings.
While precise relations between these  couplings 
 and the underlying QCD parameters are unknown,
an estimate of the magnitude of the couplings
can be obtained by applying 
Georgi and Manohar's 
Naive Dimensional Analysis (NDA)\cite{GEORGI84b,FURNSTAHL97,FRIAR96b}.

The procedure is to extract from each term in the lagrangian 
the dependence on two primary 
physical scales of the effective theory: 
the pion decay constant $f_\pi \approx 94\,$MeV
and
a larger mass scale $\Lambda\approx 4\pi f_\pi/\sqrt{N_f}$ (where $N_f$ is 
the number  of light flavors) \cite{GEORGI84,GEORGI93b}.
The mass scale $\Lambda$ is associated with the 
new physics beyond the pions:
the non-Goldstone boson masses or the nucleon mass.
This mass scale ranges from the scalar mass ($\approx 500\,$MeV) to
the baryon mass ($\approx 1\,$GeV);
the value $\Lambda \approx 600\,$MeV is consistent with all existing
empirical studies \cite{FURNSTAHL97,RUSNAK97}.  
To establish the canonical normalization of the strongly interacting
fields, 
an inverse factor of $f_{\pi}$ is included 
for each field,  and an
overall factor of $f_{\pi}^{2}\Lambda^{2}$ fixes the normalization
of the lagrangian.
The physics of NDA is discussed further in Refs.~\cite{FURNSTAHL97} 
and \cite{SEROT97}.

We restrict our attention to closed-shell nuclei, so that densities
are static and three-vector currents are zero
(see Ref.~\cite{RUSNAK97} for generalized expressions).
The energy functional is therefore
an expansion in powers of the nucleon
scalar, vector, isovector-vector, tensor, and isovector-tensor densities
scaled according to NDA:
\beqa
 \rhost &\equiv& {\rhos\over f_\pi^2\Lambda}\equiv
           {\overline \psi \psi\over f_\pi^2\Lambda}\ , 
		     \label{eq:rhost}\\
 \rhoBt &\equiv & {\rhoB\over f_\pi^2\Lambda}
	\equiv{\psi^\dagger \psi\over f_\pi^2\Lambda}\ ,  
	         \label{eq:rhoBt}\\
 \rhothreet &\equiv &{\rhothree\over f_\pi^2\Lambda} 
	\equiv\frac{1}{2}\,{\psi^\dagger\tauthree \psi
		\over f_\pi^2\Lambda}\ ,
		      \label{eq:rhotaut} \\
 \tensort_i &\equiv & {\tensor_i\over f_\pi^2\Lambda} \equiv
	   {\overline \psi\sigma^{0i} \psi\over f_\pi^2\Lambda}\ ,
	   \label{eq:rhotensort}\\
 \tensorthreet{}_i &\equiv & 
	  {\tensorthree{}_i\over f_\pi^2\Lambda}
	\equiv \frac{1}{2}\,{\overline \psi \tauthree\/\sigma^{0i} \psi \over
		f_\pi^2\Lambda}\ ,  \label{eq:rhotautensort}
\eeqa
where $\psi$ denotes the nucleon field.
The functional is also organized according to an expansion in 
powers of derivatives
acting on these densities; NDA dictates that 
each gradient acting on a density is scaled by
$\Lambda$:
\beq
   \wt\grad  \equiv  {\grad\over \Lambda}\ . \label{eq:eqsix}
\eeq

We will begin by considering only the isoscalar contributions
to the energy functional, which dominate the bulk properties of
nuclei.
The functional is (following the notation of Ref.~\cite{RUSNAK97}):
\beqa
E &=&
  \int\!d^3x\,\sum_{\alpha}^{\rm occ}\,\overline \psi_\alpha(
     -i\beta\bbox{\alpha\cdot}\grad + M)\psi_\alpha 
 \nonumber\\
 & &  \quad\null
        + f_\pi^2\Lambda^2\int\!d^3x\,\bigg\{
         \wt\kappa_2\scalarN^{\smk 2}
        -\wt\kappa_d  ({\wt\grad}\scalarN)^2  
   +\wt\kappa_3\rhost^{\smk 3}
   +\wt\kappa_4\rhost^{\smk 4}
   +\wt\eta_1 \rhoBt^{\smk 2}\rhost
   +\wt\eta_2 \rhoBt^{\smk 2}\rhost^{\smk 2}
   \nonumber\\
 & &  \qquad\qquad\qquad\qquad\null+\wt\zeta_2\rhoBt^{\smk 2}
    -\wt\zeta_d ({\wt\grad}\rhoBt)^2 
    +\wt\zeta_4 \rhoBt^{\smk 4}
    -\wt\alpha_1 \scalarN  ({\wt\grad}\scalarN)^2
    -\wt\alpha_2 \scalarN  ({\wt\grad}\rhoBt)^2
     \nonumber\\
 & &\qquad\qquad\qquad\qquad\null 
 + \mbox{isovector, tensor, and electromagnetic terms}
	\nonumber\\
 & &\qquad\qquad\qquad\qquad\null 
 + \mbox{higher-order terms}    \bigg\} 
            \ .
 \label{eq:functional} 
\eeqa
Here the sum is over occupied single-particle states $\psi_\alpha$,
which may be different for protons and neutrons. 
Isovector and tensor terms will be discussed below.
The electromagnetic energies follow closely the results from
semi-empirical mass formulas \cite{SEEGER75}.

Naive dimensional analysis 
provides an organizational principle that directly translates
into numerical estimates 
and an ordering of the terms in Eq.~(\ref{eq:functional}).
For example, each additional power of $\rhos$ is accompanied by
a factor of $\fpi^2\Lambda$.
The ratios of scalar and vector densities to this factor at nuclear matter
equilibrium density are between 1/4 and 1/7 \cite{FRIAR96b}, which 
serves as an expansion parameter. 
Similarly, one can anticipate good convergence
for gradients of the densities, since the relevant scale for
derivatives in finite nuclei should be roughly the nuclear surface
thickness $\sigma$, and so the dimensionless expansion parameter
is $1/\Lambda\sigma \le 1/5$.
The expansion is useful because the coefficients have been shown
empirically to be ``natural,'' that is, of order unity \cite{SEROT97,RUSNAK97}.

Since naturalness is valid,
a semi-quantitative estimate of the contribution of
each term in Eq.~(\ref{eq:functional}) to the energy per particle
can be made.
In previous work, 
estimates were made for the contributions at the equilibrium point
of symmetric nuclear matter by using the equilibrium density
for both the baryon and scalar densities \cite{FURNSTAHL96a,RUSNAK97}.
The estimates were compared to individual contributions directly
evaluated from Eq.~(\ref{eq:functional}),
but many terms do not contribute in nuclear matter.

Here we extend this analysis to finite nuclei with $Z$ protons,
$N$ neutrons, and $A \equiv N+Z$.
This requires
estimates for the isoscalar scalar and vector densities in the
finite system as well as for isovector, tensor, and gradient densities.
Since we must allow for variations in natural coefficients as well
as uncertainties in the scale $\Lambda$,
it is sufficient to use simple local-density approximations.

The scalar and vector densities differ by less than 10\%
in ordinary nuclei, so we can use the average baryon
density for both:
\beq
    \rhost,\rhoBt \longrightarrow
	  { \langle \rhoB \rangle \over \fpi^2\Lambda }
	  \ ,  \label{eq:rhoBavg}
\eeq
where the average value of $f(x)$ is calculated as
\beq
	 \langle f(x) \rangle \equiv {1\over A}  \int\!d^3x\, f(x)\,\rhoB(x)
	 \ .  \label{eq:averages} 
\eeq
Like the surface thickness, the average nuclear
density $\langle \rhoB \rangle$ is essentially model independent.  
We determine it for each nucleus
by using densities from one of the best-fit models (all give 
the same results to within a few percent).

\begin{figure}[p]
\begin{center}
\epsfxsize=4.1in
\epsffile{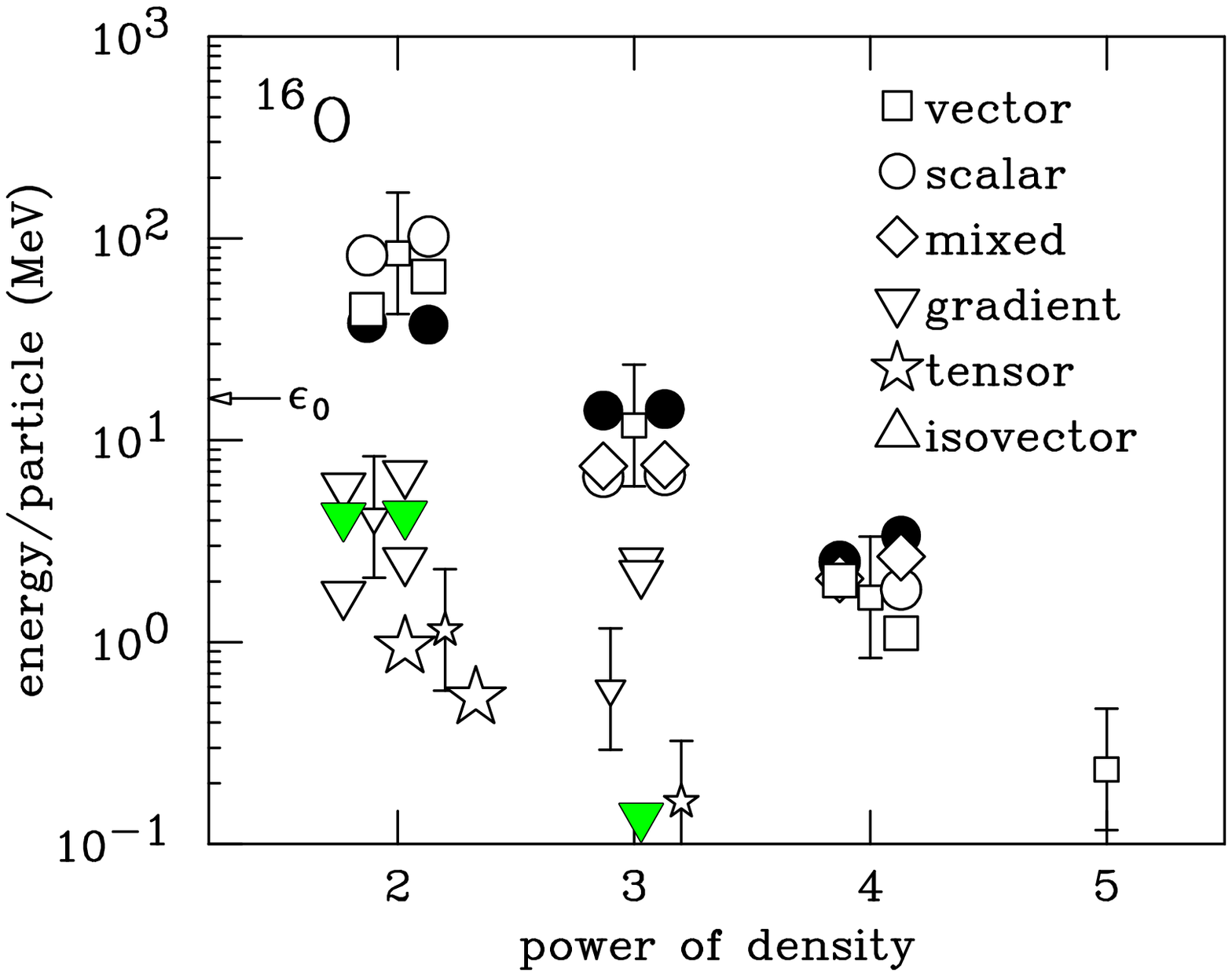}
\end{center}
\caption{Contributions to the energy per particle in ${}^{16}$O
determined by logarithmic derivatives with respect to the model
parameters (see text)
 for two RMF point-coupling models from Ref.~\protect\cite{RUSNAK97}.
 Absolute values are shown.
 The filled symbols are net values. 
The small symbols indicate estimates based on NDA
[Eq.~(\protect\ref{eq:estimate})], with the error bars
corresponding to natural coefficients from 1/2 to 2.}
\label{fig:o16energy1}

\vspace*{.1in}

\begin{center}
\epsfxsize=4.1in
\epsffile{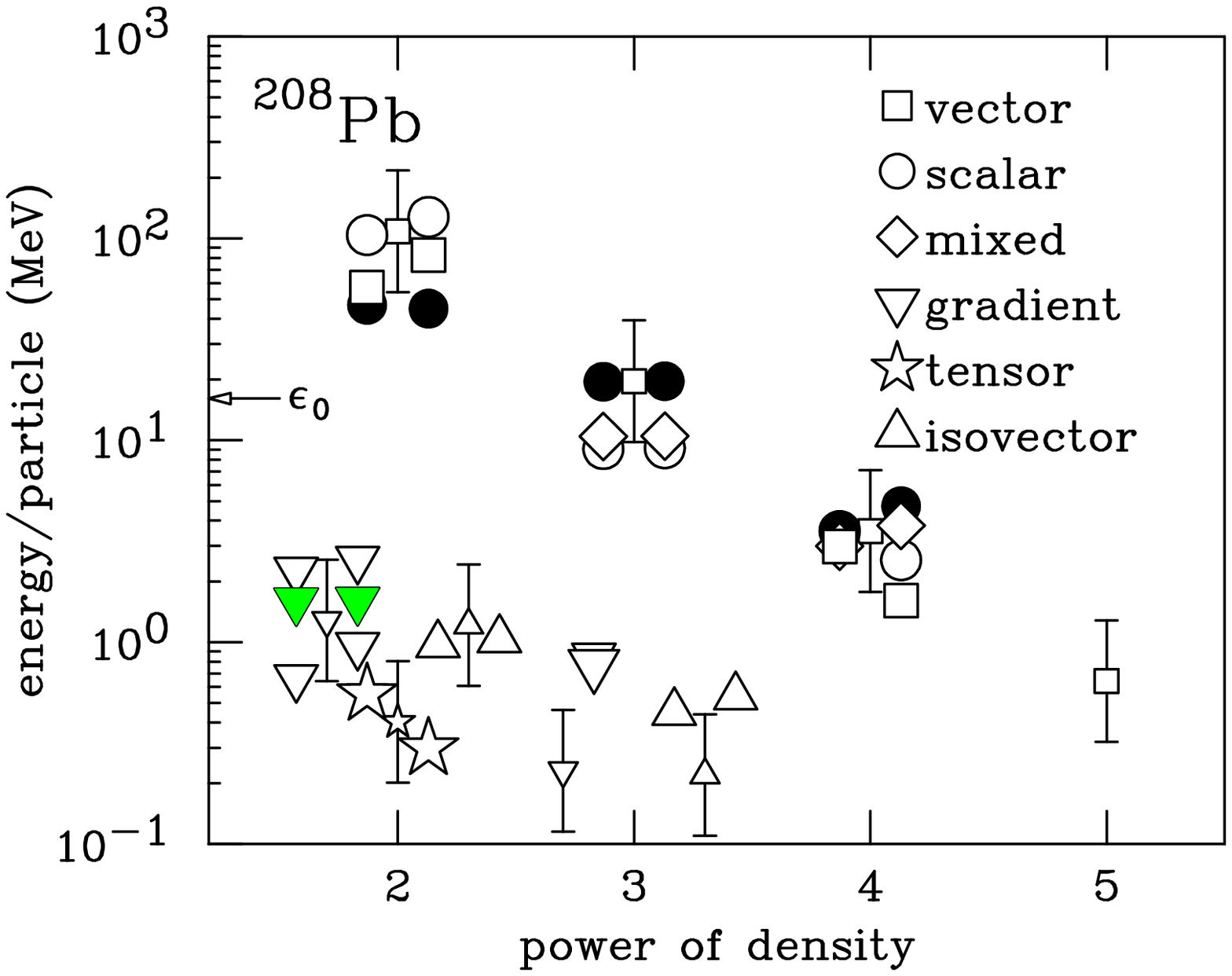}
\end{center}
\caption{Same as Fig.~\protect\ref{fig:o16energy1} for ${}^{208}$Pb.}
\label{fig:pb208energy1}
\end{figure}

A local-density 
estimate (per nucleon) for a general isoscalar term in Eq.~(\ref{eq:functional})
with dimensionless coefficient $\beta$ is made according to
\beq
   {1\over A}\fpi^2\Lambda^2 \int\! d^3x\, \beta\, (\rhost)^i (\rhoBt)^j
      \approx \beta \Lambda 
	 \left( { \langle \rhoB \rangle \over \fpi^2\Lambda } \right)^{i+j-1}
     \ ,  \label{eq:estimate}
\eeq
and we take $1/2 < \beta < 2$ as a reasonable range of natural coefficients. 
These estimates with the associated error bars from the variation
in $\beta$ are shown as small squares
in Fig.~\ref{fig:o16energy1} for ${}^{16}$O
and in Fig.~\ref{fig:pb208energy1} for ${}^{208}$Pb.
The magnitudes of energy contributions in Eq.~(\ref{eq:functional}) 
from two representative
RMF point-coupling models from Ref.~\cite{RUSNAK97} are 
shown as larger unfilled symbols (one model on each side of the error bars).
These models provide very accurate predictions of bulk 
nuclear properties \cite{FURNSTAHL96a}.
The energy contributions are determined for each nucleus by making multiple
runs while varying each parameter slightly around its optimized
value, which enables us to deduce the logarithmic derivative
with respect to each parameter.
The filled symbols denote the sum of the values for each power of the density.
The binding energy per nucleon in nuclear matter is denoted with
$\epsilon_0$.

The representative RMF models validate the isoscalar estimates from 
Eq.~(\ref{eq:estimate}), and 
the resulting hierarchy of isoscalar contributions is quite clear.
The question is:  How far down in the hierarchy can we reliably
determine contributions and their associated parameters?
To address this question, we can consider a series of models
truncated at different levels in the hierarchy.
As additional terms are added, the reproduction of experimental
observables should improve.
When improvement stops,  
contributions beyond this point are unresolved and irrelevant.

To make a meaningful analysis, we must decide on the
most relevant observables.
 The Density Functional Theory (DFT) 
perspective motivates the observables that we 
expect to calculate {\it reliably\/}  with a RMF model \cite{KOHN65}.
These include the total binding energy, density distributions
at low momentum transfer (including rms radii), the chemical
potential, and splittings between quasi-particle levels near the
Fermi surface.
We construct a measure of the accuracy of the fit to these observables,
which is based on squared deviations between predictions and data
for various observables but is {\it not\/} 
a statistical $\chi^2$.
Instead, we apply relative weightings based on our DFT expectations,
with the total 
energy weighted most heavily.
Properties of five doubly magic nuclei were used 
(see Refs.~\cite{FURNSTAHL97} and
\cite{RUSNAK97} for details).

\begin{figure}[t]
\begin{center}
\epsfxsize=3.5in
\epsffile{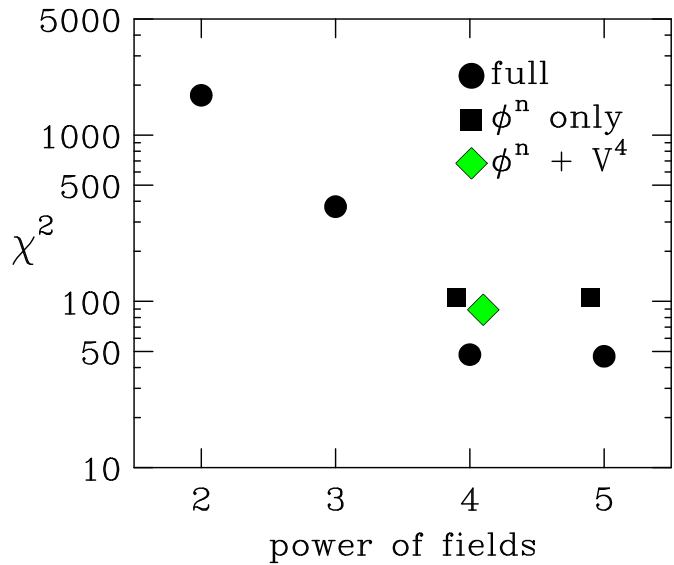}
\end{center}
\caption{Figure of merit (``$\chi^2$'') values for RMF meson models
\protect\cite{FURNSTAHL97,RUSNAK97}, ordered according to the
truncation level of the energy functional.}
\label{fig:chisq}
\end{figure}

In Fig.~\ref{fig:chisq}, the impact of different RMF meson
model truncations
is shown by plotting this figure of merit (``$\chi^2$'')
against the maximum power
of fields in a truncated energy functional.
We have performed this test with both point-coupling and meson models,
with similar conclusions.
We focus here on the meson model results because we are able to directly
compare them to standard phenomenological models.
 (See Ref.~\cite{RUSNAK97} for the point-coupling analog.)
The ``full'' models (which include all nonredundant terms at a given order) 
show that one needs to go to the fourth power
of fields to get the best fits, but going  further yields no
improvement.
Analogous behavior is found for RMF point-coupling models with
powers of densities replacing powers of fields, although the
quality of the fit is much better when the truncation is at the cubic
level
 \cite{RUSNAK97}.
Thus contributions to the energy/particle at the level of
1\,MeV or so are at the limit of resolution.
Fifth-order isoscalar contributions to the energy/particle,
which are predicted to be less than 1~MeV, 
are simply not determined by the optimization \cite{RUSNAK97}.

\begin{figure}[p]
\begin{center}
\epsfxsize=4.0in
\epsffile{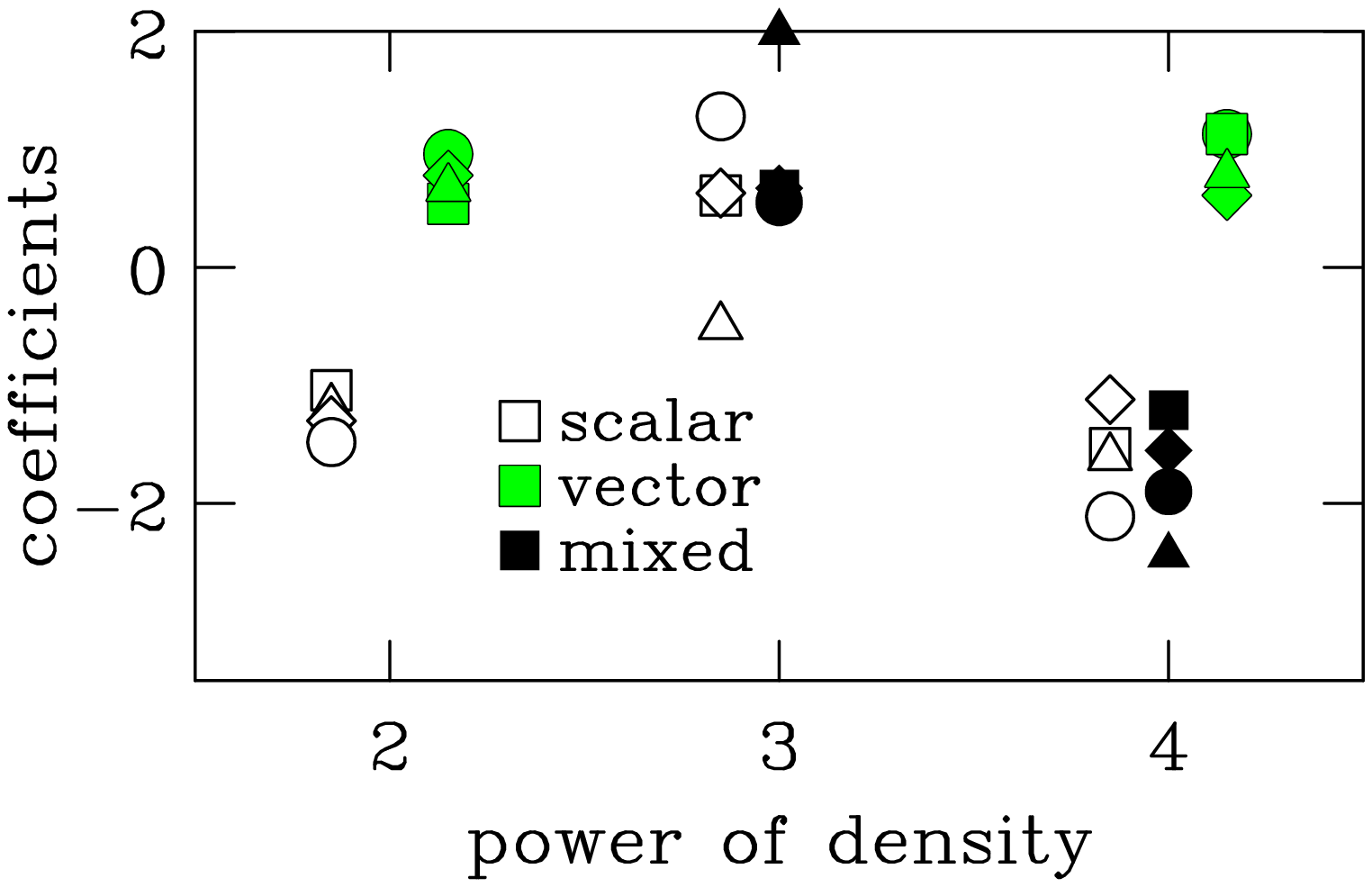}
\end{center}
\caption{Coefficients for four accurately fit RMF point-coupling
models from Ref.~\protect\cite{RUSNAK97}.  
Each model is represented by a different shape,
and the shading shows the type of term (scalar, vector, or mixed).}
\label{fig:coeffspc}
%
\vfill
%
\begin{center}
\epsfxsize=4.0in
\epsffile{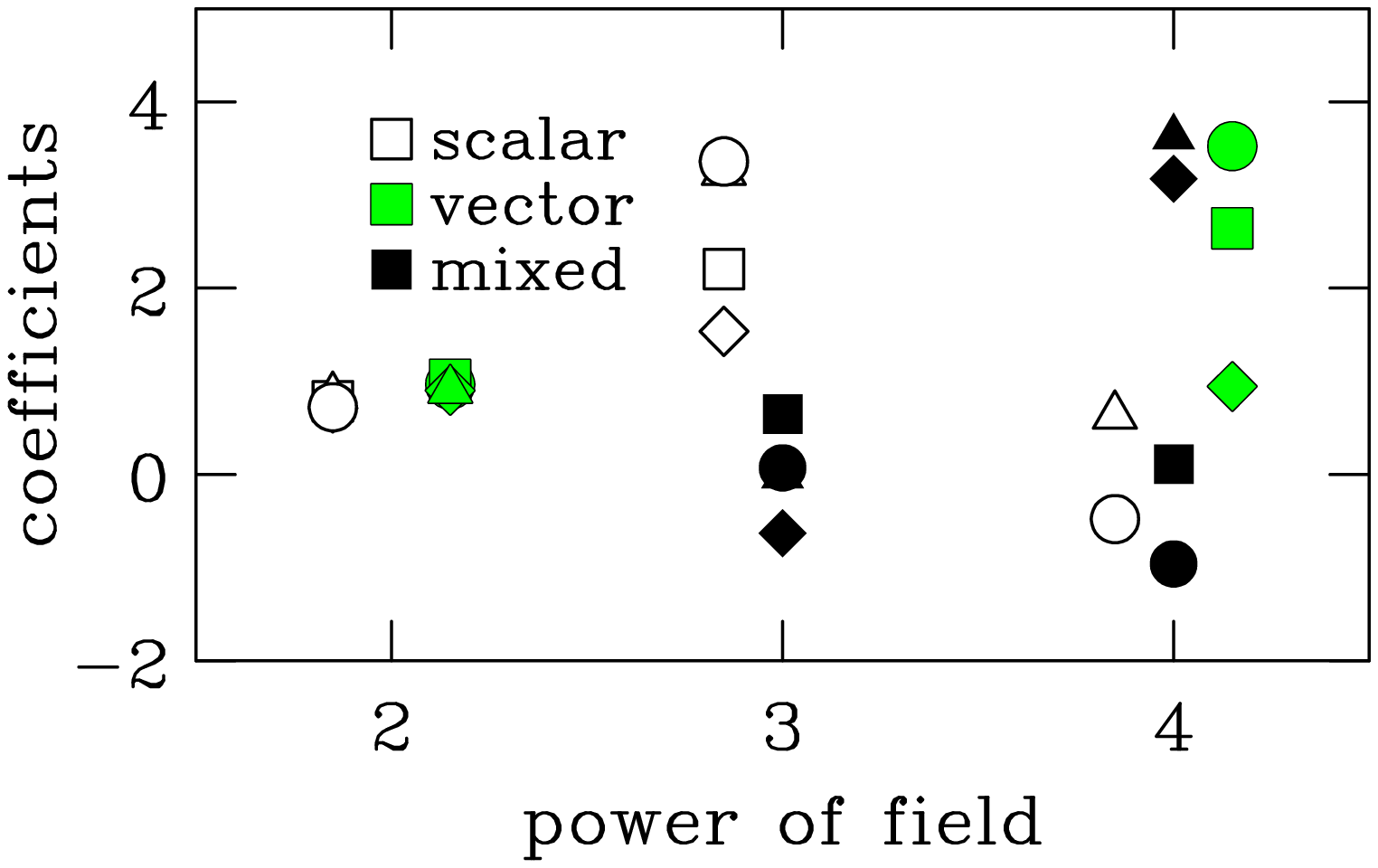}
\end{center}
\caption{Coefficients for four accurately fit RMF meson
models from Refs.~\protect\cite{FURNSTAHL97} and \protect\cite{RUSNAKTHESIS}.  
Each model is represented by a different shape,
and the shading shows the type of term (scalar, vector, or mixed).
Several parameter values are slightly beyond the graph limits and
so do not appear.}
\label{fig:coeffsvmd}
\end{figure}

The ``$\phi^n$ only'' results in Fig.~\ref{fig:chisq}, 
which include only scalar fields  
for $n>2$, show that nearly optimal results can
be obtained with just a subset
of terms at each order.  
This explains the successes of the most widely used RMF models, 
which add only $\phi^3$ and $\phi^4$ terms to the original Walecka
model lagrangian.
{\em Since the coefficients
are underdetermined by the data\/},
it is not surprising that
the systematic optimization of more complete RMF models yields
multiple EFT parameter sets with accurate 
reproductions of nuclear properties \cite{FURNSTAHL97,RUSNAK97}. 
The variation 
of coefficient values in Eq.~(\ref{eq:functional}) 
provides a measure of
how well the parameters are actually determined by the data.
In Fig.~\ref{fig:coeffspc}, the seven
coefficients of isoscalar, non-gradient terms 
from  four point-coupling models
are plotted.
The analogous results for coefficients in several meson models
is shown in Fig.~\ref{fig:coeffsvmd}. 
We note that all coefficients are natural, {\it i.e.\/}, order unity.
However, the spread in coefficient values is significant,
particularly for the meson models, and does not correspond well
to the power-counting order.
We conclude that different linear combinations of the coefficients
must be considered to draw reliable conclusions about how many
are determined by the data.

\begin{figure}
\begin{center}
\epsfxsize=3.7in
\epsffile{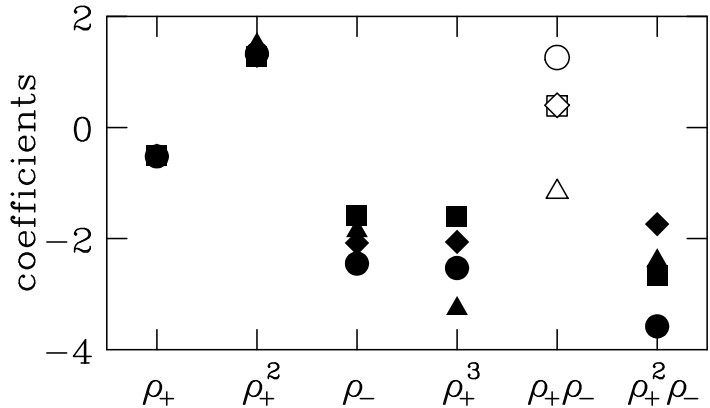}
\end{center}
\caption{Improved coefficients for the same four models as
in Fig.~\ref{fig:coeffspc}.
The ``order'' is determined by counting powers of
$\rhoplus$ and $\rhominus$.}
\label{fig:optimalpc}
\end{figure}

Can we find a more systematic power counting scheme?
The similar size of the scalar density $\rhos$ and
the vector density $\rhoB$ suggests that we
count instead powers of $\rho_+ \equiv (\rhos+\rhoB)/2$ and
$\rho_- \equiv (\rhos-\rhoB)/2$.
The corresponding  ``improved'' coefficients are listed in
Table~\ref{tab:improved} [see Eq.~(\ref{eq:functional})].
The spread in these coefficients
for four RMF point-coupling models from Ref.~\cite{RUSNAK97} 
are shown in Fig.~\ref{fig:optimalpc}.
The terms are organized according to the power of $\rho_+$
and $\rho_-$, with $\rho_-$ behaving as $\rho_+^{8/3}$ \cite{RUSNAK97}.

The leading orders are very well determined, with a 
systematic increase in uncertainty.
Even the sign is undetermined for the parameter $\wt\Omega_4$,
which is shown with unfilled symbols,
but the next parameter ($\wt\Omega_6$) 
appears to be reasonably well determined.
Higher-order terms are undetermined from the optimizations.
Deduced values and uncertainties based on this sample of models
are given in Table~\ref{tab:improved}.
We see that of the seven isoscalar, non-gradient parameters in
Eq.~(\ref{eq:functional}), four linear combinations are clearly
determined by bulk nuclear observables, with probably a fifth combination 
as well.

\def\lsim{\lower0.6ex\vbox{\hbox{$\ \buildrel{\textstyle <}
         \over{\sim}\ $}}}

\begin{table}[t]
\caption{Improved coefficients for point-coupling RMF models
\protect\cite{RUSNAK97}.}
\begin{tabular}{cccc}
coefficient & linear combination & density scaling & deduced value  \\
\hline\\[-10pt]
  $\wt\Omega_1$ & $\wt\kappa_2+\wt\zeta_2$ & $\rhoplus$ & $-0.51\pm 0.01$\\
 $\wt\Omega_3$ & $\wt\kappa_3+\wt\eta_1$ & $\rhoplus^2$ & $+1.3\pm 0.1$ \\
 $\wt\Omega_2$ & $\wt\kappa_2-\wt\zeta_2$ & $\rhominus$ & $-2.0\pm 0.4$ \\
 $\wt\Omega_5$ & $\wt\kappa_4+\wt\zeta_4+\wt\eta_2$ & $\rhoplus^3$ & $-2.4\pm 0.7$ \\
 $\wt\Omega_4$ & $\wt\kappa_3-\wt\eta_1/3$ & $\rhoplus\rhominus$ & $+0.2\pm 1.0$ \\
 $\wt\Omega_6$ & $\wt\kappa_4-\wt\zeta_4$ & $\rhoplus^2\rhominus$ & $-2.6\pm 0.8$ \\
 $\wt\Omega_7$ & $\wt\kappa_4+\wt\zeta_4-\wt\eta_22/3$ &
         $\rhoplus\rhominus^2$  &  \\[2pt]
\end{tabular}
\label{tab:improved}

\end{table}

Now we return to consider a more complete version of Eq.~(\ref{eq:functional})
by including the leading isovector and tensor terms:
\beqa
E &=&
  \int\!d^3x\,\sum_{\alpha}^{\rm occ}\,\overline \psi_\alpha(
     -i\beta\bbox{\alpha\cdot}\grad + M)\psi_\alpha 
 \nonumber\\
 & &  \quad\null
        + f_\pi^2\Lambda^2\int\!d^3x\,\bigg\{
         \wt\kappa_2\scalarN^{\smk 2}
        -\wt\kappa_d  ({\wt\grad}\scalarN)^2  
   +\wt\kappa_3\rhost^{\smk 3}
   +\wt\kappa_4\rhost^{\smk 4}
   +\wt\eta_1 \rhoBt^{\smk 2}\rhost
   +\wt\eta_2 \rhoBt^{\smk 2}\rhost^{\smk 2}
   \nonumber\\
 & &  \qquad\qquad\qquad\qquad\null+\wt\zeta_2\rhoBt^{\smk 2}
    -\wt\zeta_d ({\wt\grad}\rhoBt)^2 
    +\wt\zeta_4 \rhoBt^{\smk 4}
    -\wt\alpha_1 \scalarN  ({\wt\grad}\scalarN)^2
    -\wt\alpha_2 \scalarN  ({\wt\grad}\rhoBt)^2
     \nonumber\\[4pt]
 & &  \qquad\qquad\qquad\qquad\null+\wt\xi_2\rhothreet^{\smk 2}
    -{\wt\xi}_d  ({\wt\grad}\rhothreet)^2 
    +\wt\eta_\rho \rhothreet^{\smk 2}\scalarN
    +\wt f_v{\wt\grad}\rhoBt\cdot\bbox{\tensorN}
    +\wt f_\rho
	{\wt\grad}\rhothreet\cdot\bbox{\isovectorTensorN}
	\nonumber\\
 & & \qquad\qquad\qquad\qquad\null + \mbox{electromagnetic terms}
       + \mbox{higher order terms}
        \bigg\} \ .
 \label{eq:functfull} 
\eeqa
We will need estimates for isovector, gradient, and tensor densities.
An estimate for the isovector density is obtained by a simple
scaling of the local-density approximation to the baryon density:
\beq
    \rhothreet \longrightarrow
		 {(Z-N)\over 2A} { \langle \rhoB \rangle \over \fpi^2\Lambda }
	 \ ,	\label{eq:rhotauavg}
\eeq
which reproduces the direct average of isovector densities
in best-fit models to about 10\% accuracy.
We use [see Eq.~(\ref{eq:averages})]
\beq
    (\wt\grad\rhost)^2,(\wt\grad\rhoBt)^2 \longrightarrow
	  { \langle (\grad\rhoB)^2 \rangle \over \fpi^4\Lambda^4 }
	    \label{eq:gradrhoavg}
\eeq
for gradients and
\beq
  \wt{\bbox{s}} \longrightarrow
       {\langle {\bbox{s}} \rangle \over \fpi^2\Lambda}
	   \label{eq:tensor}
\eeq
for tensor terms.

Estimates for all of the terms in Eq.~(\ref{eq:functfull}) 
and the corresponding contributions from the two point-coupling models
are indicated with different small symbols in Figs.~\ref{fig:o16energy1}
($^{16}$O) and
\ref{fig:pb208energy1} ($^{208}$Pb).
Solid inverted triangles represent the net contribution of two gradient
terms at second and third order in the density expansion.
The actual contributions from the direct fits for the
two nuclei from isovector, gradient, and tensor terms exhibit significant
differences, but the estimates follow these systematics quite
reasonably. 
This enables us to predict how many parameters of each type
can be determined reliably.

The isovector terms appear only on the graphs 
for $^{208}$Pb.
The factor $(N-Z)/2A$, which is only 10\% even for Pb, severely cuts
down the sensitivity to isovector terms (especially since 
the factor must appear with even powers).
The magnitude of the
leading isovector term ($\propto \wt\xi_2\rhothreet^2\,$)
is comparable to the fourth-order isoscalar
term, which is at the limit of what can be determined reliably 
from fitting the binding energy.
Fits of the subleading  term ($\propto \wt\eta_\rho\rhost\rhothreet^2$)
are borderline unnatural, which is typical
for energy contributions that are essentially unresolved.
Furthermore, while individual values for $\wt\xi_2$
and  $\wt\eta_\rho$ vary widely between different parameter
sets, the {\it combination\/} 
$\wt\xi_2 + \wt\eta_\rho\langle\rhost\rangle$, with
$\langle\rhost\rangle \approx \langle\rhoBt\rangle \approx 0.16$ 
from averaging over nuclei from oxygen to lead,  is well determined.
We conclude that only one isovector parameter is determined by the bulk 
observables.
A term proportional to $(\rhothreet)^4$, which one might think important
for the neutron-star equation of state, is completely undetermined
by ordinary nuclei.%
\footnote{Fortunately, the condition of beta equilibrium in a neutron star
greatly restricts the role of a $(\rhothreet)^4$ term 
\protect\cite{MUELLER96}.}
Our results support previous observations in meson models that
at the mean-field level, the role of the $\rho$ meson is simply
to reproduce the correct symmetry energy  \cite{HOROWITZ83,MUELLER96}.
Including any other isovector degrees of freedom, such as an
isovector, scalar meson, is unnecessary.
However, the use of specific isovector observables, such as the difference
between 
neutron and proton distributions at low momentum transfer, may provide
more constraints on the isovector parameters.

The energy estimates for isoscalar tensor terms imply that only one
parameter, at best, can be determined.
Higher-spin terms, which will require more gradients and have smaller
average densities, are not at all constrained.
The tensor terms are interesting to consider because a fraction
of the spin-orbit force can be generated by including an isoscalar
tensor coupling of the vector field to the nucleon.
At the same time, the scalar and vector fields will be reduced
(and therefore the effective nucleon mass $\Mstar$ will be increased).
Nevertheless, the spin-orbit potential arises predominantly from 
the large scalar and vector fields; attributing more than one-third
of the potential to the tensor coupling produces unrealistic
surface systematics \cite{HUA99}. 
In Ref.~\cite{FURNSTAHL98}, it was shown how to accurately relate
the size of the tensor coupling to the value of  
$\Mstar$ in nuclear matter and the size of spin-orbit splittings
in finite nuclei.
 
Finally, the gradient terms follow the same pattern in the energy:
the leading term is barely above the limit of resolution.  In fact, there
are two isoscalar 
gradient terms at leading order (scalar and vector), but if
we consider linear combinations based on $\rho_+$ and $\rho_-$
(see the improved coefficients $\wt\Delta_i$ and $\wt A_i$ in \cite{RUSNAK97}),
only their sum is well determined.  
The subleading-order contributions
appear to be highly unnatural individually, but their sum  almost
vanishes.
Thus we conclude that only one gradient parameter is determined. 
This has implications for the possibility of resolving effective
meson masses from fits of point-coupling parameters.
In particular, estimates for the scalar and vector masses squared are
$\Lambda^2 \wt\kappa_2/\wt\kappa_d$ and 
$\Lambda^2 \wt\zeta_2/\wt\zeta_d$, respectively.
Since $\wt\kappa_d$ and $\wt\zeta_d$ are not individually determined,
only the difference of the inverse squared masses can be extracted.


\section{Implications and Summary}

We have shown that EFT power counting and
naturalness lead to reliable estimates of the
energy contribution in a finite nucleus
from any term in a general RMF energy functional.
By examining energy systematics based on these estimates and the spread
among coefficient values in models fit to a given set of nuclear observables,
we can deduce how many parameters are actually constrained by
those data.
We conclude that four (possibly five) isoscalar, non-gradient parameters, 
one gradient parameter, and one isovector parameter
are well determined by the usual bulk nuclear observables
(binding energies, charge density distributions, and spin-orbit splittings
in doubly magic nuclei).
Since conventional RMF models include this distribution of parameters
(plus two additional unnecessary gradient terms),
it is not surprising that empirical surveys have found no phenomenological
reason to add additional terms \cite{REINHARD86}.

The handful of parameters that are well determined can be associated
with an equal number of nuclear properties and general features of
RMF models. 
In particular,  
\begin{enumerate}
   \item Two isoscalar, non-gradient parameters are very well determined.
   These correspond to the highly constrained values for
   the equilibrium density ($\kfermi \approx 1.30\pm 0.01\,\mbox{fm}^{-1}$) 
   and binding energy ($16.0\pm 0.1\,\mbox{MeV}$)
   of observed nuclear matter.
   \item An additional isoscalar constraint is that 
   $\Mstar \approx 0.61\pm 0.03$,  if the isoscalar tensor term is set to zero.
   This range ensures an accurate reproduction of spin-orbit splittings
   in finite nuclei.
    Small increases in $\Mstar$ without changing the splittings
	can be accommodated by including
	an isoscalar tensor term, with the trade-off given by a
	simple local-density approximation in Ref.~\cite{FURNSTAHL98}. 
   \item A fourth isoscalar constraint comes from the
   nuclear matter compressibility.  
   The constraint is much weaker, in the range of $K \approx 250 \pm 50\,$MeV.
   \item The possibility of a fifth isoscalar constraint has been
   considered by Gmuca \cite{GMUCA92}, who argued that 
   one needed separate scalar
   and vector fourth-order terms to  tune the density
   dependence of the scalar and vector parts of the baryon self-energies.
   This would correspond to constraining $\wt\Omega_6$ from
   Table~\ref{tab:improved}.  In addition, some form of isoscalar nonlinear
    vector coupling is needed to soften the high-density
   equation of state to be consistent with observed neutron star 
   masses \cite{MUELLER96}.
   \item Since only one isoscalar gradient parameter is determined,
   it is not useful to allow the scalar and vector masses 
   (or their equivalents in a point-coupling model) to
   vary independently.  Thus it is convenient to fix the vector
   mass at a natural size, such as the experimental mass for the $\omega$.
   Then a scalar mass of  $500\pm 20\,$MeV is required.
  (It may be fine tuned using the ${}^{40}$Ca rms charge radius 
  3.45\,fm as a target.)
   \item The one isovector parameter can be fixed by
     the surface-corrected 
     volume symmetry energy \cite{SEEGER75}, which falls
	 in the range $34\pm 4\,$MeV \cite{HOROWITZ81}.  
     Since no isovector gradient is determined,
	 setting the isovector vector meson mass to the experimental
	 $\rho$ meson mass is adequate.  
\end{enumerate}

To produce the best quantitative fits to nuclear properties for any given model,
the model parameters should be obtained from
an optimization procedure involving calculations of a set of finite
nuclei from oxygen to lead \cite{FURNSTAHL96a,RUSNAK97}.
However, to ensure a reasonable (if not optimal) reproduction 
of finite nuclei,
it is sufficient to reproduce the nuclear matter properties given above.
Note that {\it all\/} properties must be satisfied and the 
resulting parameters must
be natural.%
\footnote{A further caution is that there are many correlations among
these properties, so that the allowed ranges should not be considered truly
independent.}

This last point has important implications for an alternative class of
RMF investigations, which are not based  on EFT but are defined by some specific
underlying physics.
Examples include models that contain different degrees
of freedom (quarks) \cite{SAITO94,SAITO96}, 
or different types of meson--baryon
couplings (derivative couplings) \cite{ZIMANYI90,CHIAPPARINI97}, 
or explicit inclusion of chiral 
symmetry \cite{KERMAN74,BIRBRAIR82,BOGUTA83,SARKAR85,HEIDE94,CARTER96}.
(The chronology for this development can be gleaned from some recent review
articles \cite{SEROT86,SEROT92,SEROT97}, the papers cited above, and references 
therein.)
Unfortunately, in many of these studies it was believed to be sufficient to
reproduce just the nuclear matter equilibrium point, and thus the
results for finite nuclei were not as accurate as expected
from the EFT analysis \cite{FURNSTAHL93,FURNSTAHL98}.
{\em One cannot justify the underlying physics of a model
if it can reproduce only a subset of the nuclear calibration data.}

Moreover, both the input observables and the parameters are highly
correlated and noisy; therefore, it is inappropriate to use the
deduced parameters from Table~\ref{tab:improved} instead of a direct
fit to observables.
It is possible to achieve a sensible picture of nuclei by this
procedure but not a very accurate fit.

The improved coefficients from Table~\ref{tab:improved}
suggest that a nonrelativistic point-coupling EFT,
with an expansion in $\rhoB$ only,
should be consistent with similar power-counting estimates.
Indeed, the phenomenologically successful Skyrme
models are in this category, and an  
NDA study shows that they are natural \cite{FURNSTAHL97c}.
A scale of $\Lambda=600\,$MeV is consistent with the trends
in all of the relativistic and nonrelativistic models. 
A comparison of these approaches from the EFT perspective
will be given elsewhere.

The EFT/DFT perspective provides insight into which additional
observables might further constrain the parameters of a general RMF functional.
Since four or five isoscalar, non-gradient terms in the energy functional
are determined, it is unlikely that additional bulk, isoscalar
observables will provide further constraints. 
This is consistent with calculations of deformed even-even nuclei, which 
reproduce experimental systematics once the energy functional
is constrained to reproduce bulk properties of doubly magic nuclei.
On the other hand, since only one isovector parameter is fixed,
additional isovector observables could be relevant.
In particular, an accurate measurement of the neutron radius in
a heavy nucleus, which would result from proposed parity violation
experiments,  may fix an additional isovector parameter.
Additional constraints on gradient terms might be provided from data on 
ground-state currents at nonzero (but low) momentum
transfer, for example, from elastic magnetic scattering.
An assessment of these constraints within the framework described
here is in progress.

\vspace*{-2pt}

\acknowledgments

We thank H.~Mueller and N.~Tirfessa for useful comments.
This work was supported in part by the National Science Foundation
under Grant No.\  PHY--9800964 and by the U.S. Department of Energy
under Contract No.~DE-FG02-87ER40365.

\end{document}